\documentstyle[12pt,epsfig]{article}
\author{Juli\'an Candia$^{a}$ and Ezequiel V. Albano$^{b}$\\
$^a${\it Departamento de F\'{\i}sica, UNLP, 
CC67,}\\{\it 1900 La Plata, Argentina}\\
$^b${\it Instituto de Investigaciones Fisicoqu\'{\i}micas
Te\'{o}ricas y Aplicadas}\\{\it (INIFTA), UNLP, CONICET, 
Suc.4, CC16,}\\{\it
1900 La Plata, Argentina}}

\title{Order-disorder criticality, wetting, and morphological 
phase transitions in
the irreversible growth of far-from-equilibrium magnetic films}
\begin{document}
\maketitle

\begin{abstract}
An exhaustive numerical investigation of the growth of magnetic films
in confined $(d+1)$-dimensional stripped geometries ($d=1,2$) is carried 
out by means of extensive Monte Carlo simulations. Films in contact with 
a thermal bath at temperature $T$, are grown by adding spins having 
two possible orientations and considering ferromagnetic (nearest-neighbor)
interactions. At low temperatures, thin films of thickness $L$ are 
constituted by a sequence of well-ordered domains of average 
length $l_D\gg L$. These domains have opposite magnetization. So, the
films exhibit ``spontaneous magnetization reversal'' during the 
growth process. Such reversal occurs within a short characteristic 
length $l_R$, such that $l_D\gg l_R\sim L$. 
Furthermore, it is found that for $d=1$ the system is non-critical, 
while a continuous order-disorder phase transition at finite temperature 
takes place in the $d=2$ case. Using standard finite-size scaling 
procedures, the critical temperature and some relevant critical exponents 
are determined. Finally, the growth of magnetic films
in $(2+1)$ dimensions with competing short-range magnetic fields 
acting along the confinement walls is studied. Due to the antisymmetric 
condition considered, an interface between domains with spins having 
opposite orientation develops along the growing direction. Such an 
interface undergoes a localization-delocalization
transition that is the precursor of a wetting transition in the 
thermodynamic limit. Furthermore, the growing interface also
undergoes morphological transitions in the growth mode. A comparison between 
the well-studied equilibrium Ising model and the studied irreversible 
magnetic growth model is performed throughout. Although valuable analogies 
are encountered, it is found that the nonequilibrium nature of the latter 
introduces new and rich physical features of interest.  
\end{abstract}

\section{Introduction}

The preparation and characterization of magnetic nanowires and
films is of great interest for the development of advanced
microelectronic devices. Therefore, the study of the
behavior of magnetic materials in confined geometries,
e.g. thin films, has attracted both 
experimental \cite{gra,iron,nw1,hong,shen,tsay,nw2,w110} 
and theoretical \cite{alba1,lan,kar,ou,reis} attention.
From the theoretical point of view, most
of the work has been devoted to the study of equilibrium properties of
thin magnetic films \cite{alba1,lan,kar,ou,reis,let}
and magnetic materials, see e.g. \cite{cos,lima,mina,klo}.
In contrast, the aim of this work is to study
the properties of thin magnetic film growth under
far-from-equilibrium conditions, by means of extensive
Monte Carlo simulations.
Within this context, this work is related to many recent 
investigations concerned with irreversible growth processes.
Indeed, the study of growth systems under far-from-equilibrium
conditions is a subject that has drawn great attention 
during the last decades. Nowadays, this interdisciplinary field 
has experienced a rapid progress due to its interest in many 
subfields of physics, chemistry, and even biology, as well as by its relevance
in numerous technological applications such as the development of
nanoscale devices, polymer science, crystal and polycrystalline growth,
gelation, fracture propagation, epidemic spreading, 
colloids, etc. \cite{fam,shl1,shl2,bar,mar}. 

For the purpose of studying the properties of thin magnetic film 
growth under far-from-equilibrium conditions, 
a variant of the irreversible Eden growth model \cite{eden}, 
in which particles are replaced by spins that can 
adopt two different orientations, is investigated.
Our study is performed in confined (stripped) geometries, which resemble
recent experiments where the growth of
quasi-one-dimensional strips of Fe on a Cu(111) vicinal surface
\cite{iron} and Fe on a W(110) stepped substratum \cite{w110}
have been performed. Also, in a related context, the
study of the growth of metallic multilayers has shown a rich
variety of new physical features. Particularly, the growth
of magnetic layers of Ni and Co separated by a Cu spacer
layer has recently been studied \cite{cobre}.

The growth of magnetic films with competing short-range magnetic fields,
which account for the interaction of the growing films with the substrate, 
is also investigated in the present work. The competing situation 
considered leads to rich and complex physical phenomena that exhibit 
a delicate and subtle interplay between finite-size effects, wetting, 
and interface growth mechanisms.  
Besides the technological and scientific interest that may arise from
the nonequilibrium nature of the model investigated, as already pointed out, 
this kind of nonequilibrium wetting phenomena also appears closely related to
very interesting equilibrium wetting transitions, which have attracted so 
far considerable experimental and theoretical attention. For instance, 
surface enrichment and wetting layers 
have been observed experimentally in a great variety of systems, such as 
e.g. polymer mixtures \cite{poly1,poly2,poly3}, 
adsorption of simple gases on alkali metal surfaces \cite{alk1,alk2,alk3}, 
hydrocarbons on mica \cite{chr}, etc. From the theoretical point of view, 
the study of wetting transitions at interfaces has been carried out   
by means of different approaches, such as the mean 
field Ginzburg-Landau method \cite{parr,swi}, transfer matrix 
and Pfaffian techniques \cite{macio1,macio2}, density matrix 
renormalization group methods \cite{car}, solving the
Cahn-Hilliard equation \cite{ch}, using Molecular Dynamic 
simulations \cite{md}, solving self-consistent field equations
\cite{scft}, and by means of extensive Monte Carlo 
simulations \cite{alba1,bind1,kur,mamu,kur1}. 

Finally, it should also be remarked that, although the 
discussion is presented here in terms of
a magnetic language, the relevant physical concepts could 
be extended to other systems such as fluids, polymers, and binary mixtures.  

This manuscript is organized as follows: 
in Section 2 details on the model and the simulation method are given,
Section 3 is devoted to the order-disorder critical behavior of 
magnetic Eden films, Section 4 deals with the study of interfacial phase
transitions that arise when competing short-range magnetic fields are
considered, while the conclusions are finally stated in Section 5. 

\section{The model and the simulation method}

In the classical Eden model \cite{eden} on the square lattice,
the growth process starts by adding particles to the
immediate neighborhood (the perimeter) of a seed particle.
Subsequently, particles are
stuck at random to perimeter sites. This growth process
leads to the formation of compact clusters with a self-affine
interface \cite{shl1,shl2,bar,mar}.
The magnetic Eden model (MEM) \cite{mem,jul1} considers an additional
degree of freedom due to the spin of the growing particles.
In the present work the MEM is studied in $(d + 1)-$dimensional 
rectangular geometries for $d=1,2$, as described in \cite{jul1}.
 
For the case $d=1$, the MEM is investigated on the square
lattice using a rectangular geometry $L\times M$ (with $M \gg L$).
The location of each site on the
lattice is specified through its rectangular coordinates $(i,j)$,
($1 \leq i \leq L$, $1 \leq j \leq M$).
The starting seed for the growing cluster is a column
of parallel-oriented spins placed at $j=1$
and film growth takes place
along the positive longitudinal direction (i.e. $j \geq  2$).
Periodic boundary conditions are adopted along the transverse 
direction. Then, assuming that each spin $S_{ij}$ can be 
either up or down (i.e. $S_{ij}= \pm 1$), 
clusters are grown by selectively adding
spins to perimeter sites, which are defined as the
nearest-neighbor (NN) empty sites of the already occupied ones.
Considering a ferromagnetic interaction of 
strength $J>0$ between NN spins, the
energy $E$ of a given configuration of spins is given by
\begin{equation}
E = - \frac{J}{2} \sum_{\langle ij,i^{'}j^{'}\rangle} 
S_{ij}S_{i^{'}j^{'}} \ \ ,
\end{equation}
\noindent where $\langle ij,i^{'}j^{'}\rangle$
means a summation taken over all occupied NN sites.

Analogously, the MEM in $(2+1)$ dimensions is studied using a
$L\times L\times M$ rectangular geometry ($M\gg L$). 
Each site on the lattice is now identified
through the rectangular coordinates $(i,j,k)$,
($1\leq i,j\leq L$, $1\leq k\leq M$), 
and the starting seed for the growing film is taken to be
a plane of $L \times L$ parallel-oriented spins placed at $k=1$.
The energy of the spin configuration is now calculated by
extending the summation in Eq.(1) to the three coordinates $(i,j,k)$.
For further details on the MEM defined in retangular 
geometries see also \cite{jul1}.
 
The last part of this paper (see Section 4) is devoted to the study of the 
$(2+1)$-dimensional MEM with competing short-range magnetic fields
applied along one of the transverse directions. In this case, the periodic
boundary conditions along the $j-$direction are changed to open ones, 
and competing surface magnetic fields $H>0$ ($H'=-H$) acting on
the sites placed at $j=1$ ($j=L$) are considered \cite{jul2}.  
Then, the energy of a given configuration of spins in this case 
will be given by 
\begin{equation}
E = - \frac{J}{2} \left( \sum_
{\langle ijk,i^{'}j^{'}k^{'} \rangle} 
S_{ijk}S_{i^{'}j^{'}k^{'}} \right) 
- H  \left( \sum_{\langle ik, \Sigma_1 \rangle } S_{i1k} -
\sum_{\langle ik, \Sigma_L \rangle } S_{iLk}  \right)  \ \ ,
\end{equation}
\noindent where $\langle ijk,i^{'}j^{'}k^{'}\rangle$
means that the summation in the first term is taken over all 
occupied NN sites,
while $\langle ik, \Sigma_1 \rangle$,
$\langle ik, \Sigma_L \rangle$
denote summations carried over occupied sites on
the surfaces $\Sigma_1$,  $\Sigma_L$ (defined as
the $j=1$ and $j=L$ planes, respectively).
Throughout this work we set the Boltzmann constant equal
to unity ($k_{B} \equiv 1$) and consider the
absolute temperature, energy, and magnetic fields 
measured in units of the coupling constant $J$.

The growth process of a MEM film consists in adding further 
spins to the growing film taking into account the corresponding 
interaction energies. 
A spin is added to the film with a probability proportional 
to the Boltzmann factor $\exp(-\Delta E/T)$, where $\Delta E$
is the total energy change involved. At each step,
the probabilities of adding up and down
spins to a given site have to be evaluated
for all perimeter sites.
After proper normalization of the
probabilities, the growing site and the orientation of the spin
are determined through standard Monte Carlo techniques.
Although both the interaction energy (as given either by
Eq.(1) or Eq.(2)) and the
Boltzmann probability distribution considered for the MEM are
similar to those used for the Ising model \cite{alba1},
it must be stressed that these two models
operate under extremely different conditions, namely the MEM
describes the irreversible
growth of a magnetic material and the Ising model is suitable for
the study of a magnetic system under equilibrium conditions.
In the MEM, the position and orientation of all deposited
spins remain fixed.
During the growth process, the system develops a rough growing interface
and evolves mainly along the longitudinal direction. Some lattice sites 
can remain empty even well within the system's bulk, but, since at each 
growth step all perimeter sites are candidates for becoming occupied,
these holes are gradually filled. 
Hence, far behind the active growing
interface, the system is compact and frozen.    
When the growing interface is
close to reaching the limit of the sample, 
the relevant properties of the irreversibly frozen
cluster's bulk (in the region where the growing process 
has definitively stopped)
are computed, the useless frozen bulk is thereafter erased, and
finally the growing interface is shifted 
toward the lowest possible longitudinal coordinate.
Hence, repeatedly applying this procedure, 
the growth process is not limited by the lattice length $M$.
In the present work clusters having up to $10^{9}$ 
spins have typically been grown.

\section{Order-disorder critical behavior of magnetic Eden films} 

Magnetic Eden films grown on a stripped geometry of
finite linear dimension $L$ at low temperatures
show an intriguing behavior that we call
spontaneous magnetization reversal. 
In fact, we have observed that long clusters are constituted
by a sequence of well ordered magnetic domains
of average length $l_{D} \gg L$.
Let $l_{R}$ be the characteristic length for the occurrence
of the spontaneous magnetization reversal.
Since $l_{R}\sim  L$, we then conclude that the phenomenon
has two characteristic length scales, namely $l_{D}$ and $l_{R}$,
such that $l_{D} \gg l_{R} \sim L$. Hence, the spontaneous 
magnetization reversal is essentially due to the small size of 
the thin film and it becomes irrelevant in the thermodynamic limit.  
Figure 1 shows a snapshot configuration of the $(1+1)-$dimensional MEM
where this phenomenon can be recognized. 
Here the reversal occurring between a domain of spins up 
(on the left side) and other one constituted by spins down (on the
right), as well as the interface between both domains, can be clearly
observed. The magnetization change occurs quite abruptly within
the characteristic length $l_{R}\sim  L$.
\begin{figure}
\centerline{{\epsfxsize=3.8in \epsfysize=2.0in \epsfbox{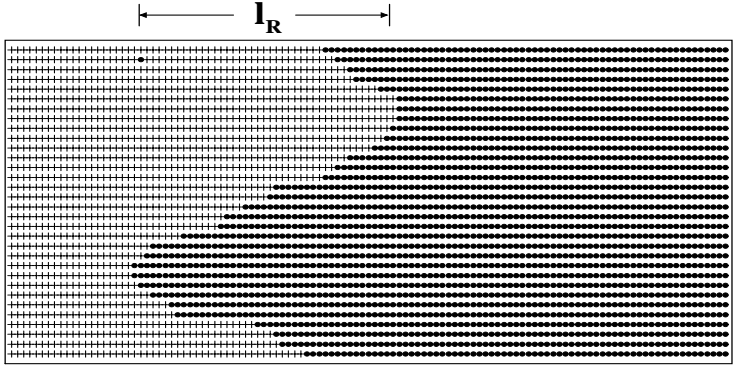}}}
\caption{Spontaneous magnetization reversal observed for $L=32$ and $T=0.26$
in the $(1+1)-$dimensional magnetic film.
The snapshot configuration shows the collective orientation 
change: the left (right) domain is constituted by up (down)
oriented spins. The snapshot corresponds to the bulk of the
sample and the growing interface is not shown.
The characteristic length for the occurrence of the
magnetization reversal, $l_R$, is of the order of the lattice width,
as marked in the figure.}
\label{fig1}
\end{figure}
In ordinary thermally driven phase transitions, the system
changes from a disordered state at high temperatures to a
spontaneously ordered state at temperatures below some critical
value $T_c$, where a second-order phase transition takes place.
Regarding the Ising model,
one has that, in the absence of an external magnetic
field ($H=0$), the low-temperature ordered phase is a state with
non-vanishing spontaneous magnetization  ($ \pm M_{sp}$).
This spontaneous symmetry breaking is possible
in the thermodynamic limit only. In fact,
it is found that the magnetization $M$ of a finite sample formed
by $N$ particles, defined by
\begin{equation}
M(T,H=0) = \frac{1}{N} \sum_{i = 1}^{N} S_{i}(T,H=0)  \ \ ,
\end{equation}
can pass with a finite probability from a value near
$+M_{sp}$ to another
near $-M_{sp}$, as well as in the opposite direction. Consequently,
the magnetization of a finite system, averaged over a
sufficiently large observation time, 
vanishes irrespective of the temperature. That is,
the equation $M(T,H=0) \approx 0$ holds if the observation time ($t_{obs}$)
becomes larger than the ergodic time ($t_{erg}$), which is defined as
the time needed to observe the system passing from $\pm M_{sp}$ 
to $\mp M_{sp}$. 
Since Monte Carlo simulations are restricted
to finite samples, the standard procedure to avoid the problems
treated in the foregoing discussion is to consider the
absolute magnetization as an order parameter \cite{kuku}.
Turning back to the MEM, we find  that the phenomenon of magnetization
reversal also causes the
magnetization of the whole cluster to vanish even for very low (but non-zero)
temperatures, provided that the film's total length $l_F$ (which plays the 
role of  $t_{obs}$) is much larger than $l_D$ (which plays the role
of $t_{erg}$). Therefore, as in the case of the Ising model \cite{kuku},
in order to overcome shortcomings derived from the
finite-size nature of Monte Carlo
simulations we have measured the mean absolute column magnetization, given by
\begin{equation}
|m(j,L,T)| = \frac{1}{L} |\sum_{i = 1}^{L} S_{ij}| \ \ .
\end{equation}

It is found that $|m(j;L,T)|$ exhibits
a transient growing period with a characteristic length of order $L$, 
followed by the attainment of a stationary regime, which is 
independent of the orientation of the seed.

\begin{figure}
\centerline{{\epsfxsize=3.8in \epsfysize=2.5in \epsfbox{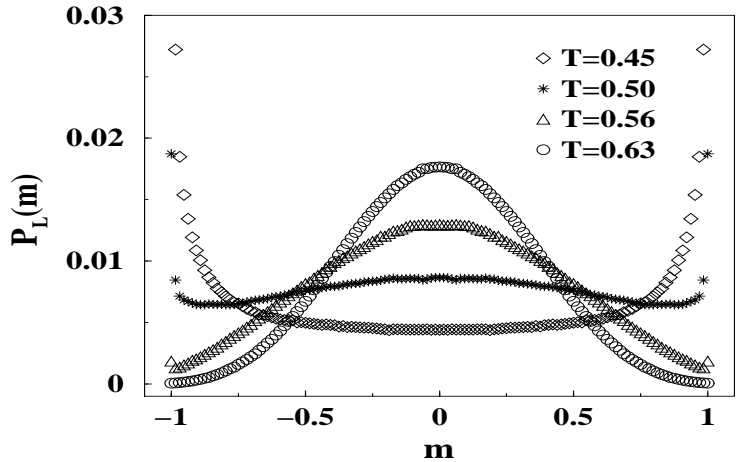}}}
\caption{Data corresponding to the $(1 + 1)-$dimensional MEM:
plots of the probability distribution of the mean
column magnetization $P_{L}(m)$ versus $m$ for the fixed lattice
width $L=128$ and different temperatures, as indicated in the figure.
The sharp peaks at $m = \pm 1$ for $T=0.45$ have been
truncated, in order to allow a detailed observation of the plots
corresponding to higher temperatures. This behavior resembles that 
of the one-dimensional Ising model.}
\label{fig2}
\end{figure}

The mean column magnetization given by Eq.(4)
is a fluctuating quantity that can assume $L+1$ values.
Then, for given values of both $L$ and $T$,
the probability distribution of the mean column
magnetization $(P_L(m))$
can straightforwardly be evaluated, since it represents the
normalized histogram of $m$ taken over a sufficiently large number
of columns in the stationary region \cite{alfa,beta,gamma}.
In the thermodynamic limit ($L\rightarrow\infty$)
the probability distribution $(P_{\infty}(m))$
of the order parameter of an equilibrium system
at criticality is universal (up to re-scaling of the
order parameter) and thus it contains
very useful and interesting information
on the universality class of the system \cite{pd1,pd2,pd3}.
For example, $P_L(m)$ contains information about all
momenta of the order parameter $m$, including
universal ratios such as the Binder cumulant \cite{pd1}.

Figure 2 shows the thermal dependence of $P_L(m)$ versus $m$,
as obtained for the $(1 + 1)-$dimensional MEM.
We can observe that at high temperatures $P_L(m)$ is a Gaussian
centered at $m=0$. As the temperature gets lowered, the distribution
broadens and develops two peaks at $m=\pm 1$.
Further decreasing the temperature
causes these peaks to become dominant
while the distribution turns distinctly non-Gaussian, exhibiting a
minimum just at $m=0$.
It should be pointed out that the emergence of the maxima
at $m= \pm 1$ is quite abrupt.
This behavior reminds us the order parameter probability
distribution characteristic of the one-dimensional Ising model.
In fact, for the well studied $d-$dimensional Ising model \cite{gamma,lali},
we know that for $T>T_c$, $P_L(M)$\cite{note} is a
Gaussian centered at $M=0$, given by
\begin{equation}
P_L(M) \propto \exp \left({-M^{2}L^{d}} \over {2T \chi} \right) \ \ ,
\end{equation}
where the susceptibility $\chi$ is related to order parameter fluctuations by
\begin{equation}
\chi = \frac {L^{d}}{T} \left(\langle M^{2} \rangle - \langle M \rangle^{2} \right) \ \ . 
\end{equation} 
Decreasing the temperature, the order parameter probability
distribution broadens, it becomes non-Gaussian, and near $T_c$ it
splits into two peaks that get the more separated the lower the
temperature. For $T<T_c$ and linear dimensions $L$ much larger
than the correlation length $\xi$ of order parameter fluctuations,
one may approximate $P_L(M)$ near the peaks by a double-Gaussian
distribution, i.e.
\begin{equation}
P_L(M)\propto\exp\left({-(M-M_{sp})^2L^d}\over{2T\chi}\right) +
\exp \left({-(M+M_{sp})^2L^d}\over{2T\chi}\right)\ \ ,
\end{equation}
where $M_{sp}$ is the spontaneous magnetization, while
the susceptibility $\chi$ is now given by
\begin{equation}
\chi=\frac{L^d}{T}\left(\langle M^2\rangle -\langle |M|\rangle^2 \right) \ \ .
\end{equation}
\noindent From Eq.(5) it turns out that
the Gaussian squared width $\sigma^{2}$
associated with high temperature distributions is very close to
the 2nd moment of the order parameter, i.e.
\begin{equation}
\sigma^2\approx\langle M^2\rangle\ \ .
\end{equation}
\begin{figure}
\centerline{{\epsfxsize=3.8in \epsfysize=2.5in \epsfbox{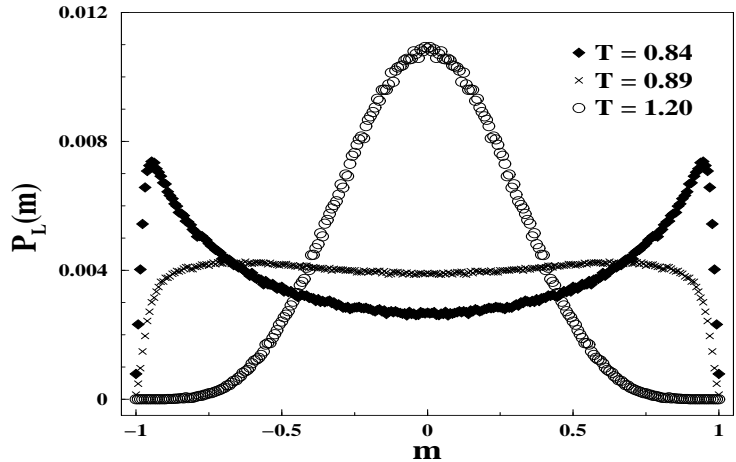}}}
\caption{Data corresponding to the $(2+1)-$dimensional MEM:
plots of the probability distribution
$P_L(m)$ versus $m$ for the fixed lattice
size $L=16$ and different temperatures, as indicated in
the figure. The occurrence of two maxima located at
$m=\pm M_{sp}$ (for a given value of $M_{sp}$
such that $0 < M_{sp} < 1$)
is the hallmark of a thermal continuous phase transition
that takes place at a finite critical temperature.}
\label{fig3}
\end{figure}
It should be noticed that this equation is a straightforward
consequence of the Gaussian shape of the order parameter probability
distribution and, thus, it holds for the MEM as well.
From the well known one-dimensional exact solution for a
chain of $L$ spins \cite{delta} one can establish the relationship
\begin{equation}
\chi = \frac {1}{T} \exp (2/T) \ \ ;
\end{equation}
\noindent then, Eqs.(6) and (10) lead us to
\begin{equation}
\langle M^{2} \rangle = \frac {1}{L} \exp (2/T) \ \ 
\end{equation} 
(where it has been taken into account that
$\langle M \rangle = 0$ due to finite-size effects, irrespective
of temperature).
From Eqs.(9) and (11) we can see that the high-temperature 
Gaussian probability distribution broadens exponentially
as $T$ gets lowered, until it develops delta-like peaks at $M= \pm 1$
as a consequence
of a boundary effect on the widely extended distribution.
It should be noted that for $d \geq 2$ this phenomenon
is prevented by the finite
critical temperature which splits the Gaussian, as implied by Eq.(7).

Figure 3 shows the thermal evolution of the probability
distribution as obtained for the $(2+1)-$dimensional MEM.
Notice that now $m$ is defined by an average over transverse planes
constituted by $L\times L$ spins, analogously to Eq.(4). Hence, 
$m$ takes now $L^2+1$ possible values.   
For high temperatures, the probability distribution
corresponds to a Gaussian centered at $m=0$. At lower
temperatures we observe the onset of two
maxima located at $m=\pm M_{sp}$ $(0<M_{sp}<1)$,
which become sharper and approach $m=\pm 1$ as $T$ is gradually decreased.
These low-temperature probability distributions
clearly reflect the occurrence of the
magnetization reversal effect already discussed for
the case of $(1+1)-$dimensional magnetic films.

Figure 4 shows the location of the maximum of the probability distribution
as a function of temperature for both
$(d+1)$-dimensional MEM models (with  $d=1,2$),
where only maxima located at $m\geq 0$ are considered,
since the distributions are symmetric around $m=0$.
After inspection of figure 4,
it becomes apparent the different qualitative behavior of both systems.
Indeed, while for the $d=2$ case we observe a smooth
transition from the $m_{max}=0$ value characteristic of high temperatures
to nonzero $m_{max}$ values that correspond to lower temperatures,
the curve obtained for $d=1$ shows instead a Heaviside-like jump.
In contrast to the $(1+1)-$dimensional case, the behavior exhibited by the
$(2+1)$-dimensional MEM
(e.g. as displayed by figures 3 and 4) is the signature
of a thermal continuous phase transition that takes place at
a finite critical temperature. 

\begin{figure}
\centerline{{\epsfxsize=3.8in \epsfysize=2.1in \epsfbox{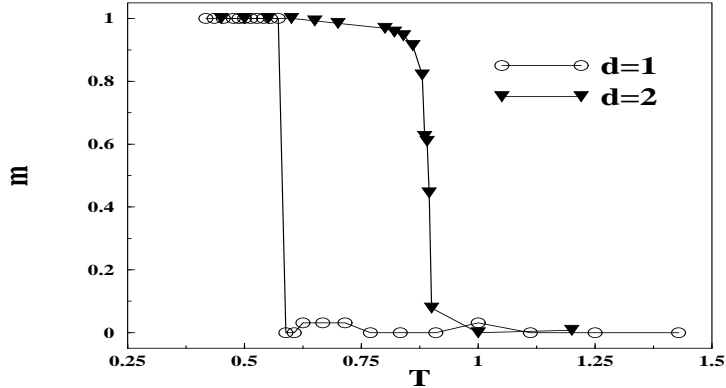}}}
\caption{Plots showing
the location of the maximum of the probability distribution
as a function of temperature for both
$(d+1)-$dimensional MEM models ($d=1,2$). The lines are guides to the eye.
The smooth transition for d=2 is the signature
of a thermal continuous phase transition occurring at
a finite critical temperature.} 
\label{fig4}
\end{figure}

From the finite-size scaling theory, developed for the treatment of
finite-size effects at criticality and under 
equilibrium conditions \cite{barba,priv}, 
it is well known that if a thermally driven
phase transition occurs at a temperature $T_c>0$ in the
thermodynamic limit, then in a confined geometry 
of linear dimension $L$ this
transition becomes smeared out over the temperature region
$\Delta T(L)$ around a shifted effective transition temperature
$T_{c}(L)$, which are related to $L$ through phenomenological exponents. 
Indeed, it is found that
\begin{equation} 
\Delta T(L)\propto L^{-\theta}
\end{equation} 
and 
\begin{equation}              
|T_{c}(L)-T_{c}|\propto  L^{-\lambda} \ \  ,
\end{equation}         
where the rounding and shift exponents are given by
$\theta=\lambda=\nu^{-1}$, respectively, and where
$\nu$ is the exponent that characterizes the divergence
of the correlation length at criticality.

Furthermore, from well-established finite-size scaling relations,
the following Ans\"atze hold just at criticality:
\begin{equation} 
\langle |m(L,T=T_c)|\rangle\propto L^{-\beta/\nu} 
\end{equation}
and
\begin{equation}                  
\chi_{max}(L)\propto L^{\gamma/\nu} \ \ ,   
\end{equation} 
where $\beta$ and $\gamma$ are the order
parameter and the susceptibility critical exponents,
respectively. Note that $\chi_{max}(L)$, as given by Eq.(15),
refers to the maximum of $\chi(L,T)$ as a function of $T$ for a
fixed lattice size $L$.
 
In order to describe quantitatively 
the critical behavior of the MEM in $(2+1)$-dimensions,
we may test the validity of the scaling relations given in Eqs.(12)-(15). 
As in the case of equilibrium systems, 
in the present case various ``effective'' $L$-dependent critical 
temperatures can also be defined. In particular, we will define $T_{c1}(L)$ 
as the value that corresponds to $\langle |m|\rangle=0.5$ for fixed $L$,
and $T_{c2}(L)$ as the one corresponding to the maximum of the
susceptibility for a given $L$, assuming that the susceptibility
is related to order parameter fluctuations in the same manner
as for equilibrium systems (as given by Eqs.(6) and (8)).
Then, $T_c$ can be obtained from
plots of $T_{cn}(L)$ versus $L^{-1}$  (for $n=1,2$),
as is shown in figure 5.
Following this procedure we find that both $T_{c1}(L)$ and $T_{c2}(L)$
extrapolate (approximately) to the same value, allowing us 
to evaluate the critical temperature
$T_c=0.69\pm 0.01$ in the thermodynamic limit. 

\begin{figure}
\centerline{{\epsfxsize=3.8in \epsfysize=2.3in \epsfbox{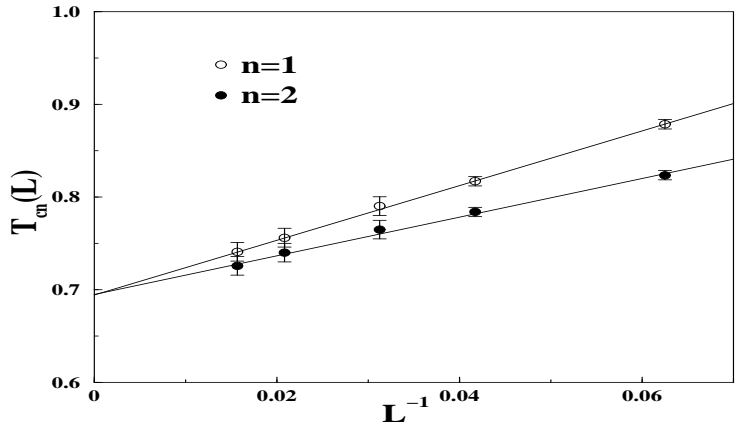}}}
\caption {Plots of the effective finite-size critical temperatures
$T_{cn}(L)$ versus $L^{-1}$ (for $n=1,2$)
corresponding to the $(2+1)-$dimensional magnetic film.
$T_{c1}(L)$ is defined
as the value that corresponds to $\langle |m|\rangle = 0.5$,
while $T_{c2}(L)$ is the temperature that corresponds to
the maximum of the susceptibility.
The solid lines show the linear extrapolations that meet at the
critical point given by $T_c=0.69\pm 0.01$.}
\label{fig5}
\end{figure}

After determining $T_c$,
the correlation length exponent $\nu$
can be evaluated by means of Eq.(13), making the replacement
$\lambda=1/\nu$. Indeed, taking $T_c$ 
at the mean, maximum and minimum
values allowed by the error bars, we obtain six log-log plots
of $|T_{cn}(L)-T_{c}|$ versus $L$ for $n=1,2$.
The slope of each of these plots, not shown here for the sake of space,
yields a value for $\nu$. The obtained values are:
$$
\nu = 1.08\ \ (T_c=0.68),\ \ \ \
\nu = 1.00\ \ (T_c=0.69),\ \
$$
\begin{equation} 
\nu = 0.88\ \ (T_c=0.70)\ \ \ \ \rm{for} \ \ n=1,
\end{equation}
and
$$
\nu = 1.20\ \ (T_c=0.68),\ \ \ \
\nu = 1.08\ \ (T_c=0.69),\ \ \ \
$$
\begin{equation}
\nu = 0.95\ \ (T_c=0.70)\ \ \ \ \rm{for} \ \ n=2.
\end{equation} 
Thus our estimate is given by $\nu=1.04 \pm0.16$,
where the error bars reflect the
error derived from the evaluation of $T_c$, as well as the
statistical error.

Studying the susceptibility $\chi$ as a function of the temperature for
several different lattice sizes, it is found that $\chi$
exhibits a peak, which becomes sharper and shifts toward lower
temperatures as $L$ is increased. Hence, Eq.(15) can be used to evaluate 
$\gamma/\nu$ from the slope of a log-log plot of $\chi_{max}$ versus $L$,
as figure 6 shows. The linear fit yields $\gamma/\nu = 2.02 \pm 0.04$.
Using this value and the value formerly obtained for $\nu$ we thus
determine $\gamma= 2.10 \pm 0.36$.

\begin{figure}
\centerline{{\epsfxsize=3.8in \epsfysize=2.5in \epsfbox{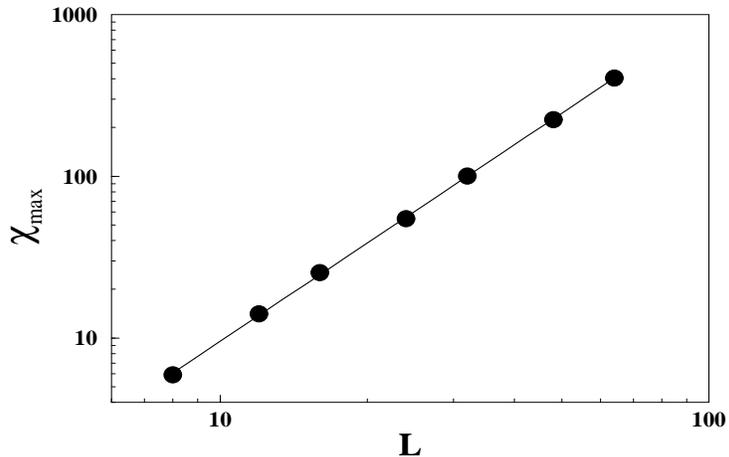}}}
\caption {Log-log plot of $\chi_{max}$ versus $L$.
The linear fit (solid line) yields $\gamma/\nu=2.02\pm 0.04$.}
\label{fig6}
\end{figure}

\begin{figure}
\centerline{{\epsfxsize=3.8in \epsfysize=2.5in \epsfbox{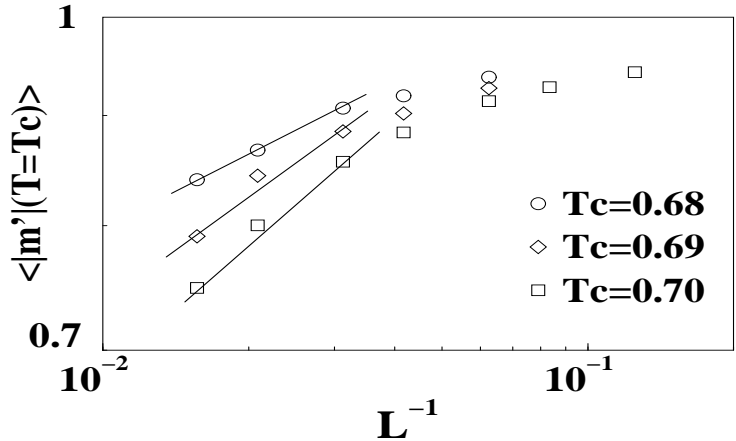}}}
\caption{Log-log plots of average magnetization measured at the
critical point $\langle m(T=T_c)\rangle$ versus $L^{-1}$ as obtained
for the mean, maximum and minimum allowed values of $T_c$.
The linear fits (solid lines) yield an estimate 
$\beta/\nu=0.16\pm 0.05$.}
\label{fig7}
\end{figure}

Figure 7 shows log-log plots of $\langle |m|\rangle(T=T_c)$ versus $L$
for the mean, maximum and minimum allowed values of $T_c$.
Considering only the larger lattices, the linear fits to the data  
according to Eq.(14) yield the following estimates: $\beta/\nu=0.11$, 
$\beta/\nu=0.16$ and  $\beta/\nu=0.19$. We then assume the value 
$\beta/\nu=0.15\pm 0.04$, where the error bars reflect the
error derived from the evaluation of $T_c$, as well as the
statistical error. From this value and the value formerly obtained 
for $\nu$ we thus determine $\beta=0.16\pm 0.05$. 

In this manner, we conclude that magnetic Eden films grown on stripped 
$(d+1)$-dimensional geometries are non-critical for $d=1$, in the
sense that the ordered phase is trivially found only at $T =0$. 
However, MEM films exhibit a continuous order-disorder phase transition at the
critical temperature $T_c = 0.69 \pm 0.01 $ for $d=2$. Furthermore,
the critical exponents of the MEM in $(2+1)$ dimensions,
as obtained using a finite-size scaling analysis, are:
$\nu = 1.04 \pm 0.16 $, $\gamma = 2.10 \pm 0.36$, and
$\beta = 0.16 \pm 0.05$. 

\section{Study of the $(2+1)$-dimensional MEM with competing short-range 
magnetic fields: Wetting and morphological phase transitions}

In this section, we will study the interfacial phase transitions 
that arise in the $(2+1)$-dimensional MEM, when competing short-range 
magnetic fields applied along one of the transverse directions are 
considered. As described in Section II, we will assume competing 
surface magnetic fields $H>0$ ($H'=-H$) acting on the surfaces
$j=1$ and $j=L$. Hence, the energy associated to a given spin 
configuration acquires an additional term due to the interaction of 
the surface spins with the applied magnetic fields (see Eq.(2)).  

As shown in figure 8, magnetic Eden films that grow in a confined 
geometry with competing surface fields exhibit a very rich phase 
diagram, which is composed of eight regions. These regions are delimited 
by several distinct, well-defined transition curves. As will be shown 
below, the bulk order-disorder (finite-size) critical point $T_c(L)$, 
the Ising-like quasi-wetting transition
curve $T_w(L,H)$, and two morphological transitions associated to the 
curvature of the growing interface (namely, from convex to 
non-defined to concave), 
can be quantitatively located. Moreover, in order to gain some 
insight into the physics involved in this complex phase diagram, 
some typical snapshot configurations characteristic 
of the various different growth regimes are obtained (see figure 9) 
and discussed. Finally, the phase diagram in the thermodynamic 
limit will be drawn (see inset of figure 8) by extrapolating 
finite-size results (see figure 10).  

The ($L-$dependent) bulk order-disorder critical temperature can be 
identified with the peak of the susceptibility at zero surface field. 
For $L=12$, the critical point so defined is $T_c(L=12) = 0.84$,
and is shown in figure 8 by a vertical straight line.
So, the left (right) hand side part of
the phase diagram corresponds to the ordered (disordered) growth
regime that involves Regions $I,II,III,IV$, and $A$ (Regions $V,VI$, and $B$). 

\begin{figure}
\centerline{\epsfxsize=3.8in \epsfysize=2.7in \epsfbox{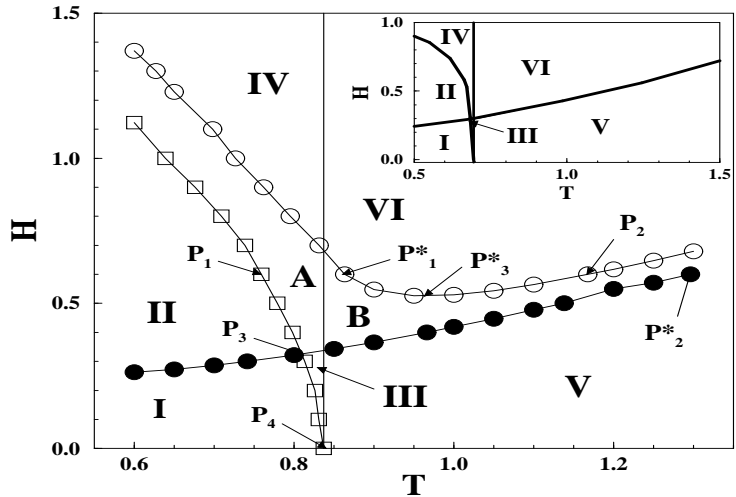}}
\caption{$H-T$ phase diagram corresponding to a lattice of size $L=12$. 
The vertical straight line at $T_c(L) = 0.84$ corresponds 
to the $L-$dependent critical temperature, which
separates the low-temperature ordered phase
from the high-temperature disordered phase.
Open (filled) circles refer to the transition between non-defined and 
concave (convex) growth regimes, and squares stand for the Ising-like 
localization-delocalization transition curve.
Eight different regions are distinguished, 
as indicated in the figure. 
Also indicated are seven representative points that are 
discussed in the text. The inset shows the phase diagram
corresponding to the thermodynamic limit composed of
six different regions.}
\label{fig8}
\end{figure} 

Using a standard procedure \cite{alba1}, the localization-delocalization 
transition curve (on the $H-T$ plane) corresponding to the up-down 
interface running along the walls can be computed, considering that 
a point with coordinates $(H_w,T_w)$ on this curve maximizes $\chi(H,T)$. 
So, the size-dependent localization-delocalization transition curve
is obtained, as shown in figure 8 (open squares).
As in the case of the Ising model, this quasi-wetting 
transition refers to a transition
between a nonwet state that corresponds to a localized 
interface bound to one of the confinement walls, 
and a wet state associated with a 
delocalized domain interface centered between roughly equal domains of
up and down spins.
The localization-delocalization transition 
in a confined system is indeed the precursor of the
true wetting transition, which occurs in the thermodynamic 
limit \cite{alba1,swi,bind1}.
In fact, there is observed a finite jump in the wetting layer thickness
that takes place as a result of the finite size of the system.
As the lattice size is increased, the magnitude of the jump grows
and diverges in the $L\rightarrow\infty$ limit, as expected 
for a continuous wetting transition. 

Since the MEM is a nonequilibrium kinetic growth model, 
it also allows the identification of another kind of phase transition, 
namely a morphological transition associated with the curvature of 
the growing interface of the system. 
To avoid confusion, we would like to remark that the 
term {\it interface} is used here for the 
transverse interface between occupied and empty lattice sites, 
while it was used above for the longitudinal
interface between up and down spin domains. 
To explore this phenomenon quantitatively, the behavior
of the contact angles between the growth interface and 
the confinement walls (as functions of temperature and magnetic field) have
to be investigated.
Clearly, two different contact
angles should be defined in order to locate this transition,
namely $\theta_D$ for the angle corresponding to the 
dominant spin cluster, and $\theta_{ND}$ for the one that 
corresponds to the non-dominant spin cluster.
Both contact angles can straightforwardly be determined by measuring
the location of the growth interface averaged over a
sufficiently long growing time.  
In this way,  three  different  growth  regimes  can  be  distinguished: 
 {\it (i)} the concave growth regime that occurs when the system 
partially wets the walls on both sides (i.e. for 
$\theta_D, \theta_{ND} < \frac{\pi}{2}$),
{\it (ii)} the convex growth regime that 
occurs for $\theta_D, \theta_{ND} > \frac{\pi}{2}$, 
and {\it (iii)} the regime of non-defined
curvature that occurs otherwise. The corresponding transition curves
obtained for confined magnetic Eden films are shown on the $H-T$ phase diagram 
of figure 8, where open (filled) circles refer to the transition 
between non-defined and  concave (convex) growth regimes. 

As anticipated above, we will now introduce and 
discuss some characteristic snapshot pictures, 
in order to provide qualitative explanations 
that account for the different growth regimes observed.
Let us begin with Region $I$ (see figure 8), that corresponds 
to the Ising-like nonwet state and the convex growth regime. 
In this region, the temperature is low and the
system grows in an ordered state, i.e. the dominant spin domain prevails and
the deposited particles tend to have their spins all pointing in 
the same direction. Small clusters with the opposite orientation 
may appear preferably on the surface where the non-dominant 
orientation field is applied. These ``drops'' might 
grow and drive a magnetization reversal, thus changing the sign 
of the dominant domain. Indeed, the formation of sequences of 
well-ordered domains are characteristic of the ordered phase of 
confined (finite-size) spin systems
such as the Ising magnet \cite{alba1}. 
Due to the open boundary conditions, 
perimeter sites at the confinement walls 
experience a missing neighbor effect, that is, the number
of NN sites is lower than for the case of perimeter sites
on the bulk. Since $H$ is too weak to compensate this effect, 
the system grows preferentially along the center of the sample 
as compared to the walls, and the
resulting growing interface exhibits a convex shape.
So, Region $I$ corresponds to the Ising-like nonwet
state and the convex growth regime.
A typical snapshot configuration characteristic of Region $I$ is 
shown in figure 9(a).

\begin{figure}
\centerline{{\epsfysize=4.2in \epsfysize=2.7in \epsfbox{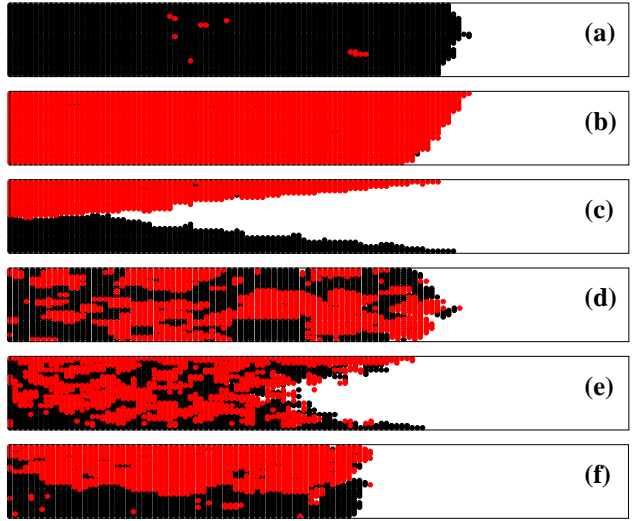}}}
\caption{Snapshot pictures showing a longitudinal slice 
given by a fixed value of the transverse coordinate $i$. 
Grey (black) points correspond to spins up (down). 
The surface field on the upper (lower) confinement
wall is positive (negative). 
The snapshots correspond to a lattice size $L=32$ and 
several different values of temperature and surface fields: 
(a)$H=0.05$, $T=0.6$; (b)$H=0.5$, $T=0.55$;
(c)$H=1.4$, $T=0.6$; (d)$H=0.1$, $T=1.0$; (e)$H=1.6$, $T=1.4$;
and (f) $H = 0.20$, $T = 0.82$.}
\label{fig9}
\end{figure} 

Let us now consider an increase in the surface magnetic fields, 
such as the system may be driven into Region $II$ (see figure 8).
Since the temperature is kept low, the system is still in its ordered phase and
neighboring spins grow preferably parallel-oriented. The surface fields
in this region are stronger and thus capable of compensating the missing
NN sites on the surfaces. But, since the fields on both surfaces have opposite
signs, it is found that, on the one hand, the field that 
has the same orientation as the dominant spin cluster favors the 
growth of surface spins, while on the other hand,
the sites on the surface with opposite field have a lower probability to
be chosen during the growth process. Hence, 
the contact angle corresponding to
the dominant spin cluster is then $\theta_D < \frac{\pi}{2} $, while the
non-dominant is $\theta_{ND} > \frac{\pi}{2} $. 
Thus, on the disfavored side the growing interface becomes pinned
and the curvature of the growing interface is not defined. 
Figure 9(b) shows a typical snapshot corresponding to Region $II$.

Keeping $H$ fixed within Region $II$ but increasing the temperature, 
thermal noise will enable the formation of drops on the disfavored side 
that eventually may nucleate into larger clusters as the temperature 
is increased even further. This process may lead to the emergence of 
an up-down interface, separating oppositely oriented domains, running in the
longitudinal direction (i.e. parallel to the walls). Since sites along
the up-down interface are surrounded by oppositely oriented NN spins, 
they have a low growing probability. 
So, in this case the system grows preferably along the confinement 
walls and the growing interface is concave (figure 9(c)). 
Then, as the temperature is
increased, the system crosses to Region $A$ (see figure 8) and 
the onset of two competitive growth regimes is observed, 
namely: {\it (i)} one  
exhibiting a non-defined growing curvature that appears when 
a dominant spin orientation is present, as in the case shown in figure 9(b); 
{\it (ii)} another that appears
when an up-down interface is established and the system
has a concave growth interface, as is shown in figure 9(c). 

On further increasing the temperature and for large enough fields, 
the formation of a stable longitudinal up-down interface
that pushes back the growing interface is observed. So,
the system adopts the concave growth regime (see figure 9(c) corresponding to
Region $IV$ in figure 8). Increasing the temperature beyond $T_c(L)$,
a transition from a low-temperature ordered 
state (Region $IV$) to a high-temperature disordered state
(Region $VI$, see figure 9(e)), both within the concave 
growth regime, is observed. Analogously, for small enough fields, 
a temperature increase drives the system from the ordered convex growth
regime (Region $I$) to the disordered convex growth regime (Region $V$,
see figure 9(d)). As shown in figure 8, there is also an 
intermediate fluctuating state (Region $B$) between Regions $V$ and $VI$, 
characterized by the competition between
the disordered convex growth regime and the disordered concave one.  

Finally, a quite unstable and small region (Region $III$ in figure 8) that
exhibits the interplay among the growth regimes of the contiguous
regions, can also be identified. Since the width of Region $III$
is of the order of the rounding observed in $T_c(L)$, large
fluctuations between ordered and disordered states are observed,
as well as from growth regimes of non-defined curvature to  
convex ones. However, figure 9(f) shows a snapshot configuration 
that is the fingerprint of Region $III$, that may prevail in the
thermodynamic limit, namely a well defined spin up-down interface 
with an almost flat growing interface. 

Let us now extrapolate our results to
show that the rich variety of phenomena
found in a confined geometry is still present in the 
thermodynamic limit ($L \rightarrow  \infty$),
leading to the phase diagram shown in the inset of figure 8.
As clearly seen by comparison with the finite-size results, 
the crossover Regions $A$ and $B$ collapse in this limit,
so only the six regions that correspond to well identified
growth regimes (as illustrated by the snapshot configurations of figure 9)
appear to remain.
  
In order to illustrate the extrapolation procedure, 
the following seven representative points of the
finite-size phase diagram are discussed in detail: 
{\it (i)} the points labeled $P_1$, $P_1^*$, $P_2$, 
and $P_2^*$, that correspond to the
intersections of the $H = 0.6$ line with the various transition
curves shown in figure 8, and {\it (ii)} the points labeled
$P_3$, $P_3^*$, and $P_4$, that refer to 
the intersection point between Regions 
$I, II, III$, and $A$, the minimum of the limiting curve between 
Regions $IV$-$VI$ and $A$-$B$, and 
the zero-field transition point, respectively.

\begin{figure}
\centerline{{\epsfxsize=3.8in \epsfysize=2.5in \epsfbox{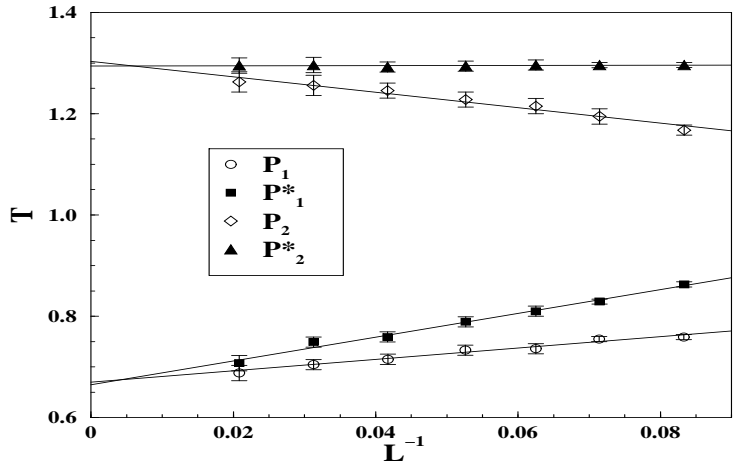}}}
\caption{Plots of $T$ versus $L^{-1}$ for $12 \leq L \leq 48$,
corresponding to the points $P_1,P_1^*,P_2$, and $P_2^*$, 
all of them with $H = 0.6$. 
The fits to the data (solid lines) show 
that, within error bars, $P_i \rightarrow P_i^*$  $(i=1,2)$ for
$L \rightarrow \infty$.}
\label{FIG. 10}
\end{figure}

Figure 10 shows plots of $T$ versus $L^{-1}$ for $12 \leq L \leq 48$ 
corresponding to the points $P_1,P_1^*,P_2,$ and $P_2^*$. 
Also shown in the figure are the fits to
the data extrapolated to $L^{-1}=0$.
The results from the extrapolations are: 
$T_1=0.67\pm0.01$, $T_1^*=0.66\pm0.01,$ and
$ T_2=1.30\pm0.02$, $T_2^*=1.29\pm0.01,$
pointing out that, within error bars, 
$P_i \rightarrow P_i^*$  $(i=1,2)$ in the $L \rightarrow \infty$
limit. Using the same procedure,
the extrapolations of $P_3$ and $P_3^*$
(not shown here) give:
$H_3=0.30\pm0.01$, $H_3^*=0.31\pm0.02,$
and $T_3=0.69\pm0.01$, $T_3^*=0.71\pm0.03$.
So, one has $P_3\rightarrow P_3^*$ for $L\rightarrow \infty$  
within error bars. Finally, the extrapolation of $P_4$ is
$T_4=T_c=0.69\pm0.01$.
  
Using the above-mentioned extrapolation procedure,
the phase diagram in the thermodynamic limit can be drawn,  
as shown in the inset of figure 8. 
By comparison with the finite-size phase diagram of
figure 8, one can note that, as anticipated, 
the crossover Regions $A$ and $B$ appear in the
phase diagram just as a consequence of the finite-size 
nature of confined geometries, since they collapse in the
$L\rightarrow\infty$ limit.
Moreover, we conjecture that Region $III$ may remain
in the thermodynamic limit. Although this (very tiny!) region corresponds to
a physically well-characterized growth regime, since
one expects that the system in this region may grow in an ordered phase
with a delocalized up-down domain interface and a convex growing interface,
statistical errors due to large fluctuations close to criticality 
hinder a more accurate location of this region. The unambiguous clarification
of our conjecture remains as an open question that will require a huge 
computational effort. 

Besides an Ising-like continuous wetting transition, coupled
morphological transitions in the growing interface, which arise
from the MEM's kinetic growth process, have also been identified.
Comparing the equilibrium wetting phase diagram of the Ising model
\cite{alba1,parr,bind1} and that of the MEM, it follows that the nonequilibrium
nature of the latter introduces new and rich  physical features of interest:
the nonwet (wet) Ising phase splits out into Regions $I$ and $II$
(Regions $III$ and $IV$), both within the ordered regime ($T<T_c$) 
but showing an additional transition in the interface growth mode. 
Also, the disordered state of the Ising system ($T>T_c$) splits
out into Regions $V$ and $VI$ exhibiting a transition in the 
interface growth mode.

\section{Conclusions}

In the present work we have studied the growth of magnetic Eden
films with ferromagnetic interactions between nearest-neighbor 
spins in a $(d+1)-$dimensional rectangular
geometry (for $d=1,2$), by means of extensive Monte Carlo simulations.
For both dimensions the phenomenon of spontaneous magnetization reversal
is observed at low temperatures. Indeed, MEM films grown
at low temperatures are constituted by a sequence of magnetic domains, 
each of them
with a well-defined magnetization, such that the magnetization of
adjacent domains is antiparallel. Furthermore, it is found that
the $(1+1)-$dimensional MEM is non-critical, while the 
$(2+1)-$dimensional MEM undergoes a thermally driven second-order
phase transition at finite temperature, which is evaluated by extrapolating
some ``effective'' $L$-dependent critical temperatures to 
the thermodynamic limit ($L\rightarrow\infty$). Using a finite-size scaling
theory, some relevant critical exponents that characterize the behavior of the
$(2+1)-$dimensional MEM at criticality are determined.
The observed behavior is reminiscent to that of the equilibrium Ising
model, although it should be stressed that the MEM
is a far-from-equilibrium growing system.

Finally, the $(2+1)-$dimensional MEM with competing surface magnetic fields,
which may account for the interaction of the growing magnetic films 
with the substrate,
is investigated. An Ising-like localization-delocalization 
wetting transition and, on the other hand, a morphological 
transition associated with
the curvature of the growing interface, are located. In this way, 
eight different regions on the $H-T$ phase diagram for a finite-size 
lattice are identified. 
Moreover, the characteristic behavior of typical growth 
processes within each region are discussed, and 
qualitative explanations that account for the observed features
are provided. Extrapolating the results obtained for various lattice sizes, 
the phase diagram corresponding to the $L\rightarrow\infty$ limit is 
also determined. It is composed of six different regions,
since two crossover regions identified in the finite-size phase diagram
appear to collapse in the thermodynamic limit. 
The phase diagram obtained shows new and rich physical features of interest, 
which arise as a consequence of the nonequilibrium nature of 
the model investigated. 

We expect that the present study will contribute to 
the fields of irreversible growth processes in confined geometries and
nonequilibrium wetting phenomena, 
and we hope that it will stimulate further experimental and theoretical 
work in these topics of widespread technological and scientific interest.

\vskip 1.0 true cm
{\bf  ACKNOWLEDGMENTS}. Work supported by
CONICET, UNLP, and ANPCyT (Argentina).


\begin{thebibliography}{99}
\bibitem{note} Notice that, according to the definitions given by
Eqs.(3) and (4), we have used $P_L(m)$ and $P_L(M)$
for the MEM and the Ising model, respectively.
\end{thebibliography}
\end{document}